\documentclass[prl,twocolumn]{revtex4}

\usepackage{graphicx}
\usepackage{amsmath, amssymb}
\usepackage{natbib}

\begin{document}

\title{Magnetoconductivity in the presence of Bychkov-Rashba spin-orbit interaction.}

\author{Alexander Punnoose}%
\email{apunnoose@wisc.edu} \affiliation{Physics Department,
University of Wisconsin at Madison, 1150 University Ave., Madison,
WI 53706}

\begin{abstract}
A closed-form analytic formula for the magnetoconductivity in the
diffusive regime  is derived in the presence of Bychkov-Rashba
spin-orbit interaction in two dimensions.
It is shown that at low fields $B\ll B_{so}$, where $B_{so}$ is the
characteristic field associated with spin precession,
D'yakonov-Perel' mechanism leads to spin relaxation, while for $B\gg
B_{so}$ spin relaxation is suppressed and the resulting spin
precession contributes a Berry phase-like spin phase to the
magnetoconductivity.
The relative simplicity of the formula  greatly facilitates data
fitting, allowing for the strength of the spin-orbit coupling to be
easily extracted.
\end{abstract}

%\pacs{}
\maketitle

%\section{Introduction}%
%
The conductivity in  classically weak magnetic fields shows
signatures of quantum interference that is considerably affected by
the presence of spin-orbit (SO) interactions. Knowing the functional
dependence of the magnetoconductivity, $\Delta\sigma(B)$, provides a
sensitive tool for the extraction of the strength of the SO
coupling. It is thus of great practical interest to obtain a simple,
analytic formula for $\Delta\sigma(B)$.

In a system with SO interactions the dominant spin relaxation
mechanism is the D'yakonov-Perel' (DP) mechanism~\cite{dp}.
It describes how momentum relaxation by impurity scattering can lead
to spin relaxation via the SO interaction. The spin relaxation rate
is given by $1/\tau_{so}\sim \Delta_{so}^2\tau/\hbar^2$, where
$\Delta_{so}$ is the spin-splitting and $\tau$ is the mean free
time.
For the special case when the SO interaction is of the
Bychkov-Rashba kind, i.e.,~\cite{vasko,rashba}
\begin{equation}
H=\frac{\vec{p}\;^2}{2m}+\alpha_{so}~\vec{\sigma}\cdot
(\hat{z}\times\vec{p}\;)~, \label{eqn:rashba}
\end{equation}
the spin splitting at the Fermi surface is
$\Delta_{so}=2m\alpha_{so}v_F$, where $\alpha_{so}$ is the SO
coupling. In this case, $1/\tau_{so}=D(2m\alpha_{so}/\hbar)^2$,
where $D=v_F^2 \tau/2$ is the diffusion constant in two dimensions
(2D) and $v_F$ is the Fermi velocity.

Earlier works treated the effect of the SO interaction exclusively
in terms of the DP relaxation~\cite{altshulerSO,HLN}, i.e., the SO
scattering rate was introduced as a cutoff $1/\tau_{so}$ in the
triplet part of the interference processes leading to the well known
anti-localization effect.
However, it became apparent on rewriting the Hamiltonian in
Eq.~(\ref{eqn:rashba}) as
%
%\begin{equation}
$H_{so}=\frac{1}{2m}\left(\vec{p}-p_{so}\vec{\tau}\ \right)^2
\label{eqn:rashbavector} $ %\end{equation}
that the SO interaction which now appears as a spin-dependent vector
potential $\vec{\tau}=\frac{1}{2}(\hat{z}\times\vec{\sigma})$ with
charge $p_{so}=2m\alpha_{so}$  could  give rise to spin-dependent
Aharonov-Bohm~\cite{gefenAB,mathurAB,gellerAB} and Berry
phase-like~\cite{gauge:AF} effects.

In 2D, the effects of various kinds of SO interactions on the
magnetoconductivity have been studied extensively in the low field
regime $l_B\gg l$~\cite{iordanskii94,knap,pikus}. In this limit the
magnetic length $l_B=\sqrt{\hbar/2eB}$ is many times larger than the
mean free path $l=v_F\tau$, and therefore the field range $B\ll
B_{tr}=\hbar/4eD\tau$ is adequately described by the diffusion
approximation~\cite{note:ballistic}.  It has been shown that an
analytical expression for $\Delta\sigma(B)$ can be derived for the
Bychkov-Rashba and the Dresselhaus kinds of SO
interactions~\cite{iordanskii94,knap}, and that no analytical
solution is possible when both terms are present
together~\cite{pikus}. These solutions are expressed in the form of
a series over effective Landau Level indices that are then summed
numerically.

In this paper, these  summations are done analytically, thereby
providing a closed-form formula for $\Delta\sigma(B)$. Although only
the Bychkov-Rashba case is worked out in detail here, the solution
is the same  in the case when only the linear Dresselhaus term is
present, i.e., when the cubic term is absent - which is the case in
low density systems. Various approximate formulas used in the
literature are derived as limiting cases of the formula derived here
and their physics highlighted.

%\section{Calculations}

As is well known, quantum correction to the conductivity arises from
the interference  of time reversed trajectories~\cite{gorkov79}. The
amplitude of this interference, $C(\textbf{r},\textbf{r}')$, is
called the Cooperon.
Using the formalism first developed in Ref.~\cite{iordanskii94} (for
a detailed review, see Ref.~\cite{knap}), it can be shown that in
the diffusive regime the Cooperon in the presence of the
Bychkov-Rashba interaction satisfies the equation: $
\mathcal{H}C(\textbf{r},\textbf{r}')=
{\delta^2(\textbf{r}-\textbf{r}')}/{2\pi\nu\tau^2}$, where
\begin{equation}
\mathcal{H}=D\left(-i\vec{\nabla}-\frac{2e}{\hbar}\textbf{A}-
\frac{p_{so}}{\hbar}\vec{\Sigma}\right)^2+\frac{1}{\tau_{\varphi}}~.\label{eqn:diffB}
\end{equation}
%\end{subequations}
%
The parameter $\nu=m/2\pi\hbar^2$ is the density of states per spin
and the rate $1/\tau_{\varphi}$ is introduced to  account for
dephasing. The spin-matrix $\vec{\Sigma}=\hat{z}\times \mathbf{S}$,
where
$\mathbf{S}=\frac{1}{2}\left(\vec{\sigma}^R+\vec{\sigma}^A\right)$
is the total  spin of the interfering waves, i.e., the retarded and
advanced waves in the particle-particle channel (Cooper
channel)~\cite{note:spinhalf}.

In terms of the dimensionless quantities,
$\tilde{\mathcal{H}}=\mathcal{H}/\hbar\omega_D$ and
$\tilde{C}=(2\pi\nu\tau^2\hbar\omega_D)C$, where the ``cyclotron"
frequency  $\omega_D=4eDB/\hbar$, the Cooperon equation reduces to
$\tilde{\mathcal{H}}\tilde{C}=\delta^2(\textbf{r})$.
In the circular gauge,
$\textbf{A}(\textbf{r})=(B/2)(\hat{z}\times\textbf{r})$, the
Hamiltonian $\tilde{\mathcal{H}}$ expressed in terms of the raising
and lowering operators reads~\cite{iordanskii94,pikus,knap}:
\begin{equation}
\tilde{\mathcal{H}}=\{aa^\dagger\}-i\sqrt{2b_{so}}%^{\frac{1}{2}}
(a^\dagger S^+ - a S^-)%\right.\nonumber\\&&\hspace{2cm}+\left.
+b_{so}(\textbf{S}^2-S_z^2)+b_{\varphi}~,\label{eqn:htilde}
\end{equation}
where, the notation $\{aa^\dagger\}\equiv
{\scriptstyle\frac{1}{2}}(aa^\dagger+a^\dagger a)$ is used. The
operators $a^\dagger$ and $a$ raise and lower the Landau level index
$n$, and $S^\pm$ raise and lower the $S_z$ values, respectively. The
dimensionless variables $b_{so}=B_{so}/B$ and
$b_{\varphi}=B_{\varphi}/B$, where
\begin{equation}
B_{so}=\frac{\hbar}{4eD\tau_{so}}=\frac{p_{so}^2}{4
e\hbar}\hspace{0.5cm} \textrm{and}\hspace{0.5cm}
B_{\varphi}=\frac{\hbar}{4eD\tau_{\varphi}}~.\label{eqn:bvalues}
\end{equation}

The term proportional to $\sqrt{b_{so}}$  in Eq.~(\ref{eqn:htilde})
mixes the Landau levels with the spin triplet states. The singlet
$S=0$ sector as seen from Eq.~(\ref{eqn:htilde}) does not mix. As a
result $\tilde{H}$ for $S=0$ reduces
to~\cite{iordanskii94,pikus,knap}
$\tilde{\mathcal{H}}_s=\{aa^\dagger\}+b_{\varphi}$, with eigenvalues
$\tilde{E}_{s}(n)=n+1/2+b_{\varphi}$. Further progress can be made
by noting that the triplet sector conveniently decomposes into
$3\times 3$ blocks~\cite{iordanskii94,knap}  around each $n$ spanned
by the three vectors $|n-S_z\rangle\otimes|S_z\rangle$ where
$S_z=-1,0,1$ for $n>0$. These blocks, labeled
$\tilde{\mathcal{H}}_n$, take the form:
\begin{equation}
\tilde{\mathcal{H}}_n=\left(\begin{array}{ccc}\epsilon_{n-1}+b_{so}&i\sqrt{2b_{so}n}&0\\
-i\sqrt{2b_{so}n}&\epsilon_n+2b_{so}&i\sqrt{2b_{so}{\scriptstyle (n+1)}}\\
0&-i\sqrt{2b_{so}{\scriptstyle(n+1)}}&\epsilon_{n+1}+b_{so}\end{array}\right)~,
\label{eqn:hmat}
\end{equation}
where, $\epsilon_n=n+1/2+b_{\varphi}$. (The $n=0$ term is treated
separately). $\tilde{\mathcal{H}}_n$ is easily diagonalized by
taking linear combinations
$|n,m\rangle=\sum_{S_z}c_{n-S_z,S_z}^m|n-S_z\rangle\otimes|S_z\rangle$.
The corresponding eigenvalues are labeled  $\tilde{E}_{t,m}(n)$,
where $m=-1,0,1$.

The  conductivity correction, $\delta\sigma(B)$, defined
as~\cite{iordanskii94}:
\begin{equation}
\delta\sigma(B)=-\frac{e^2}{2\pi
h}\sum_{\alpha\beta,n}\tilde{C}_{\alpha\beta\beta\alpha}(n)~,\label{eqn:csigma}
\end{equation}
where, $\alpha$ and $\beta$ are spin indices, can be expressed in
terms of the singlet $\tilde{E}_s(n)$ and triplet
$\tilde{E}_{t,m}(n)$ eigenvalues as~\cite{iordanskii94}:
\begin{equation}
\delta\sigma(B)=\frac{e^2}{2\pi
h}\sum_n\left[\frac{1}{\tilde{E}_{s}(n)}-\sum_{m=0,\pm
1}\frac{1}{\tilde{E}_{t,m}(n)}\right]~. \label{eqn:dsigmaenergies}
\end{equation}
To further simplify,  the sum of the inverse eigenvalues in the
triplet sector can be written as~\cite{iordanskii94}:
%
%\begin{equation}
$S_t=\sum_m{\tilde{E}^{-1}_{t,m}(n)}= \sum_i{\ \
[\tilde{\mathcal{H}}_{n}]_{ii}}/{|\tilde{\mathcal{H}}_n|}~,$
%\end{equation}
%
where $|\tilde{\mathcal{H}}_n|$ is the determinant and
$[\tilde{\mathcal{H}}_n]_{ii}$ are the minors of the diagonal
elements, giving
\begin{equation}
\!\!S_t\!\!=\!\!\!\!\sum_{n=1,\cdot\cdot}\frac{3\epsilon_n^2+4b_{so}\epsilon_n+(5b_{so}^2+4b_{so}b_{\varphi}-1)}
{\epsilon_n^3+(b_{so}^2+4b_{so}b_{\varphi}-1)\epsilon_n+2b_{so}^2(b_{so}+2b_{\varphi})}~.\label{eqn:st}
\end{equation}

Eq.~(\ref{eqn:st}) when substituted into
Eq.~(\ref{eqn:dsigmaenergies}) gives a series solution for the
magnetoconductivity (with the $n=0$ term properly included) that was
first obtained in Ref.~\cite{iordanskii94,knap}. In the following,
the sum over $n$ is performed analytically to give a closed form
expression for $\Delta\sigma(B)$. It is worth noting that one
arrives at the same equations if instead of the Bychkov-Rashba
interaction  the linear Dresselhaus term was present (see
Ref.~\cite{iordanskii94}). In the latter case the parameter $b_{so}$
takes the form~\cite{iordanskii94} $b_{so}=(\hbar/4
eD)2\Omega_1^2\tau$, where $\Omega_1=\gamma k(\langle k_z^2\rangle
-k^2/4)$. Hence, as a function of $b_{so}$ the final results derived
here are equally valid in both cases.

The crucial step to do the $n$ sum is to expand $S_t$ as:
\begin{equation}
S_t=\sum_{n=0,1,\cdot\cdot}\left[\;\sum_{s=0,\pm 1}
\frac{u_s}{\epsilon_n-v_s}\right]
+\frac{1}{\frac{1}{4}-(b_{\varphi}+b_{so})^2}~.\label{eqn:stexpand}
\end{equation}
(Note that the  $n$ sum has been extended to include the $n=0$ term.
It cancels the apparent divergence that appears in the last term in
Eq.~(\ref{eqn:stexpand}).)
The advantage of expanding $S_t$ in this way is that since
$\epsilon_n$ is linear in $n$, the sum over $n$ can be done using
the formula:
\begin{equation}
\sum_{n=0}^{\infty} \frac{e^{-\alpha_B n}}{n+z}\ \
\stackrel{{\scriptstyle \alpha_B\rightarrow 0^+}}{\!\!\!\!\approx}
-\bigl(\psi(z)+\gamma+\ln\alpha_B\bigr)~,\label{eqn:sumformula}
\end{equation}

\noindent where, $\psi(z)$ is the di-gamma function, $\gamma$ is the
Euler constant, and $\alpha_B$ is a field dependent cutoff
parameter~\cite{note:cutoffs}.
The variables $u_s$ and $v_s$ are easily obtained
as~\cite{cubicsolve}:
\begin{subequations}
\begin{eqnarray}
v_s&=&{2}\ \delta \cos\left(\theta-\frac{2\pi}{ 3}(1-s)\right)~,\label{eqn:vs}\\
u_s&=&\frac{3 v_s^2+4 b_{so} v_s+(5
b_{so}^2+4b_{so}b_{\varphi}-1)}{\prod_{s'\neq s}(v_s-v_{s'})}~,
\label{eqn:us}
\end{eqnarray}
\label{eqn:usvs}
\end{subequations}
where, $\delta$ and $\theta$ are defined as:
\begin{subequations}
\begin{eqnarray}
\delta&=&\sqrt{\frac{{1-4 b_{so}b_{\varphi}-b_{so}^2}}{3}}~,\label{eqn:delta}\\
\theta&=&\frac{1}{3}
\cos^{-1}\left(-\left(\frac{b_{so}}{\delta}\right)^3\left(1+\frac{2b_{\varphi}}{b_{so}}\right)\right)~.
\label{eqn:theta}
\end{eqnarray}
\label{eqn:deltatheta}
\end{subequations}

The magnetoconductivity is defined as
$\Delta\sigma(B)=\delta\sigma(B)-\delta\sigma(0)$. To find the
zero-field value, $\delta\sigma(0)$, the sum over $n$ in
Eq.~(\ref{eqn:st}) is replaced by an integral over  momentum $q$,
with the replacement $\omega_D n\sim Dq^2$. This gives the final
result,
\begin{widetext}
\begin{subequations}
\begin{eqnarray}
\frac{\Delta\sigma(B)}{\sigma_0}&=&\sum_{s=0,\pm 1} u_s \psi
\left(\frac{1}{2}+\frac{B_{\varphi}}{B}-v_s\right)-\psi\left(\frac{1}{2}+\frac{B_{\varphi}}{B}\right)\nonumber\\&&
\ -2\ln\left(\frac{ B_{\varphi}}{B}\right)
+\frac{4B^2}{4(B_{so}+B_{\varphi})^2-B^2}+C,\\
C&=&-2\ln\left(1+\frac{B_{so}}{B_{\varphi}}\right)-\ln\left(1+\frac{2
B_{so}}{ B_{\varphi}}\right) \nonumber\\&&\  +\frac{8}{\sqrt{7+ 16
B_{\varphi}/B_{so}}} \cos^{-1}\left(\frac{2B_{\varphi}/B_{so}-1}
{\sqrt{\left(2B_{\varphi}/B_{so}+3\right)^2-1}}\right),
\end{eqnarray}
\label{eqn:dsigma}
\end{subequations}
\end{widetext}
where, $\sigma_0=e^2/2\pi h$. The constant $C$ satisfies the
requirement $\Delta\sigma(0)=0$. Eq.~(\ref{eqn:dsigma}), combined
with Eqs.~(\ref{eqn:bvalues}), (\ref{eqn:usvs}) and
(\ref{eqn:deltatheta}), is the main result of this paper. It
provides an analytic expression for $\Delta\sigma(B)$ in the
presence of  the Bychkov-Rashba SO interaction. The formula, as
mentioned earlier, is also valid for the case of the linear
Dresselhaus interaction.
It is important to bear in mind that Eq.~(\ref{eqn:dsigma}) is only
valid in the diffusive regime $B\ll B_{tr}$. In this limit $B_{tr}$
does not appear explicitly.

The different limiting cases $(i)~B\ll B_{so}$ and $(ii)~B\gg
B_{so}$ are studied below:
For $B\ll B_{so}$, it can be shown from Eqs.~(\ref{eqn:usvs}) and
(\ref{eqn:deltatheta}),  that:
$\delta \approx (i/\sqrt{3}) b_{so}$ and $\theta\approx
\cos^{-1}(-i3\sqrt{3})$, and therefore $u_s\approx 1-i (4/\sqrt{7})
s$ and $v_s\approx b_{so}(3s^2+i s \sqrt{7}-2)/2$. (It is assumed
that $B_{so}\gg B_{\varphi}$,   valid at low temperatures). Hence,
in the low field limit, Eq.~(\ref{eqn:dsigma}) reduces to the
familiar form, first derived by Hikami, Larkin and Nagaoka
(HLN)~\cite{HLN}:
\begin{eqnarray}
\frac{\Delta\sigma(B)}{\sigma_0}\!\!&\approx&\!\!
2\psi\!\left(\frac{1}{2}+ \frac{B_{\varphi}+B_{so}}{B}\right)
-2\ln\left(\frac{B_{\varphi}+B_{so}}{B}\right)\nonumber\\
&&+\psi\!\left(\frac{1}{2}+\frac{B_{\varphi}+2 B_{so}}{B}\right)
-\ln\left(\frac{B_{\varphi}+2 B_{so}}{B}\right)\nonumber\\
&&\ \ -\psi\left(\frac{1}{2}+\frac{B_{\varphi}}{B}\right)
+\ln\left(\frac{B_{\varphi}}{B}\right),\label{eqn:smallB}
\end{eqnarray}
In the high field limit $B\gg B_{so}$,
$\delta\approx(1-2b_{so}b_{\varphi})/\sqrt{3}$, $\theta\approx
\pi/6$,  $u_s\approx 1$ and $v_s\approx (1-2b_{so}b_{\varphi})s$. Up
to a constant Eq.~(\ref{eqn:dsigma}) in this limit reduces to:
\begin{equation}
\frac{\Delta\sigma(B)}{\sigma_0}
%&\approx& \sum_{s=\pm 1}\psi
%\left(\frac{1}{2}+b_{\varphi}(1+2sb_{so})\right) -2\ln b_{\varphi}\nonumber\\
\approx\!\!\sum_{s=\pm
1}\psi\!\left(\frac{1}{2}+\frac{B_{\varphi}}{B-2s B_{so}}\right)
-2\ln\!\left(\frac{B_{\varphi}}{B}\right)~.~\hspace{0.25cm}
\label{eqn:largeB}
\end{equation}

As seen from Eqs.~(\ref{eqn:smallB}) and (\ref{eqn:largeB}) the
characteristic field scale $B_{so}$ of the SO interaction  plays
fundamentally different roles depending on the strength of $B$.
For $B\ll B_{so}$ the SO interaction as seen in
Eq.~(\ref{eqn:smallB}) leads to spin relaxation via the DP
relaxation mechanism, thus $B_{so}$ in this case appears as a
cutoff. At low temperatures, $B_{\varphi}\ll B_{so}$, the dominant
term is $\Delta\sigma(B)\approx
-\psi(1/2+B_{\varphi}/B)+\ln(B_{\varphi}/B)$ resulting in negative
magnetoconductivity~\cite{HLN}.
For $B\gg B_{so}$, on the other hand, the SO interaction appears as
a gauge field $B_{eff}=B-2sB_{so}$, akin to a Berry phase-like spin
phase (see Ref.~\cite{marcus03}), and the sign of $\Delta\sigma(B)
>0$, restoring positive magnetoconductivity. (Note that the $s=0$
sector is not affected by the SO interaction in this limit.) This
crossover from negative to positive magnetoconductivity  around
$B\approx B_{so}$  is an unambiguous signature of the presence of DP
mechanism.

To summarize, a simple closed-form analytic expression has been
derived for the magnetoconductivity in the presence of SO
interaction with linear splitting of either the Bychkov-Rashba or
the Dresselhaus kind. The relative simplicity of the formula greatly
facilitates data fitting from which the strength of the spin-orbit
coupling can be extracted~\cite{stefan06}. Note that although
$1/\tau_{so}$ depends on the transport properties of the electrons,
the scale $B_{so}$ in Eq.~(\ref{eqn:bvalues}) depends only on
$p_{so}=2m\alpha_{so}$ ~\cite{dresselhausexpt}. Hence, using
$B_{so}$ as a fitting parameter directly gives the value of
$p_{so}$.

AP greatly benefitted from discussions with A.\ M.\ Finkel'stein,
M.\ Khodas, M.\ Manfra, R.\ de\ Picciotto, S.\ Schmult and S.\ H.\
Simon.

\newpage

%\bibliography{MRSObib}

\begin{thebibliography}{20}
\bibitem[{not({\natexlab{b}})}]{note:spinhalf}
\bibinfo{note}{The `Hamiltonian' $\mathcal{H}$ in Eq.~(\ref{eqn:diffB}) is the
  spin-$1$ generalization (with net charge $2e$ and mass $1/2D$) of the
  spin-$1/2$ Hamiltonian discussed in Ref.~\cite{rashba}.}
\bibitem[{not({\natexlab{c}})}]{note:cutoffs}
\bibinfo{note}{The $B$-dependence of the cutoff $\alpha_B$ is got by comparing
  $\alpha\tau D q^2=(\alpha\tau\omega_D)n\equiv\alpha_Bn$, giving
  ${\alpha_B}=(\omega_D\tau)\alpha$, where $\alpha$ is the cutoff for $B=0$}.
\end{thebibliography}
%\bibliographystyle{apsrev}

\end{document}